\documentclass[preprint,
aps,
amsfonts,
amssymb,
nofootinbib,
tightenlines
]{revtex4}

\usepackage{amscd}
\usepackage{amsmath}
\usepackage{bm}
\usepackage[dvips]{color}                 %%% for colours in the graphs
\usepackage{graphicx}
\usepackage{eucal}
\usepackage{pslatex}

\def\){\right)} 
\def\({\left(} 
\def\]{\right]} 
\def\[{\left[}

\def\Journal#1#2#3#4{{#1} {\bf #2}, #3 (#4)}

\def\PLB{{\em Phys. Lett.} B}
\def\PRL{\em Phys. Rev. Lett.}

\def\PRC{{\em Phys. Rev.} C}

\def\PRA{{\em Phys. Rev.} A}

\newcommand{\mcal}[1]{{\mathcal #1}}

\begin{document}
\preprint{INT PUB-06 5}

\title{Universality in a 2-component Fermi System  at Finite Temperature}

\author{Gautam Rupak
%\footnote{Email: {\tt grupak@u.washington.edu}} 
}
\email{grupak@u.washington.edu}
\affiliation{Institute for Nuclear Theory\\
 University of Washington\\
Seattle, WA 98195}

\begin{abstract}
Thermodynamic properties of a fermionic system close to
the unitarity limit, where the 2-body scattering length $a$
approaches $\pm\infty$ is studied at high temperature Boltzmann
regime. For dilute systems the virial expansion
coefficients in the Boltzmann regime are expected, from general arguments,
to be \emph{universal}. 
A model independent finite temperature $T$ calculation of 
 the third virial coefficient $b_3(T)$ is presented. At the unitarity limit, 
$b_3^\infty\approx-0.48$ is an universal number. 
The energy density up to the third virial expansion 
is derived. These results are relevant for dilute neutron matter and 
could be tested in 
current atomic experiments on dilute fermionic gases near the
Feshbach resonance.  
\end{abstract}

\maketitle

%=======================================================================
%\begin{section}{Introduction}
%=======================================================================

Recent experiments in atomic traps near the Feshbach 
resonance~\cite{JEThomas,TBourdel,KDieckmann,DJin,SJochim} have 
opened opportunities to study properties of finite density systems 
that have not been explored before. Using an external magnetic field, 
it is possible to tune the 2-body $S$-wave 
scattering length $a$ essentially 
at will: it can be made arbitrarily large
($a=\pm\infty$) compared to the range of the 2-body interaction
$R$. Several groups have taken measurements near the Feshbach resonance 
in $^6$Li and $^{40}$Ka fermionic atomic gases
~\cite{JEThomas,TBourdel,KDieckmann,DJin,SJochim}. 
These experiments are important checks of theoretical tools, both new
and old, in previously untested territories and hold the prospect of
revealing many new phenomenon.
Equally important, what we 
learn from these atomic studies have universal application in many
other subfields of physics.   

At densities $n$ that are dilute compared to the range of the 
interactions ($n R^3 \ll1$), the physics should be insensitive to the
details of the interaction at the short distance scale $R$. Therefore,
 near the Feshbach resonance, as $|a|\rightarrow\infty$, where there aren't any
relevant scale left in the interaction, its properties are expected to
be \emph{universal} and not just applicable to only atomic systems.
Many physical systems in nature are close to this universal limit. 
In $^4$He atomic gases, 
the scattering length $a\sim 100 \mathring{A}$ is much
larger than the range of the interaction $R\sim 5\mathring{A}$.     
To take an example from nuclear physics, the neutron-neutron $S$-wave 
scattering length   
$a_{nn}\sim -19$ fm is much larger than the range of the interaction 
set by the pion mass $R\sim \hbar/m_\pi\sim 1.4$ fm.
The universal
properties learned in atomic experiments are applicable to
other problems in the same universality class that are  otherwise not
directly accessible such as neutrino physics in supernova which depends on
properties of dilute nuclear matter with a large resonating scattering
length~\cite{HorowitzSchwenk}. 
At the typical temperatures $T\sim 4$
MeV~\cite{Costantini:2004ry,Lunardini:2004bj} and densities $n\sim
10^{-4}$ fm$^3$~\cite{HorowitzSchwenk} 
of the neutrinosphere, it is
possible to calculate the equation of state to a
given order in $n\lambda^3$ expansion, where 
$\lambda=\sqrt{2\pi/(MT)}$ is the non-relativistic thermal wavelength
of a fermion with mass $M$. We calculate the equation of state up to
the third order in $n\lambda^3$ and 
it model independently describes dilute fermionic
systems at finite temperature with a large scattering length 
even if the microscopic physics is difficult to calculate or poorly
understood.   

In this paper, we study the properties of a spin-$\frac{1}{2}$
non-relativistic fermion at finite temperature $T$ 
such that the thermal wavelength  
$\lambda= \sqrt{2\pi/(MT)}\gg R$ and we are not sensitive to the short
distance scale $R$. However,
the temperature is large enough such that the thermal wavelength is small
compared to the inter-atomic distance  $n \lambda^3\ll 1$
and scattering length $\lambda\lesssim a$. Near the Feshbach resonance with 
$|a|\rightarrow\infty$, the hierarchy of momentum scales would be then  
$|a|\gg n^{-1/3}\gg \lambda\gg R$. The calculation is more
general and applicable for $n\lambda^3\ll 1$, $\lambda\gg R$. 

The calculation is organized as an
expansion in $n \lambda^3\ll 1$, 
the Boltzmann regime, with factors of
$|a|\sqrt{MT/(2\pi)}$ summed to all orders in perturbation. In the 
Boltzmann regime the pressure $P$ for a 2-component fermi system 
can be written in terms of the so called \emph{virial} expansion~\cite{Huang}:
\begin{align}\label{virial_exp}
\frac{P}{T}&= \frac{2}{\lambda^3}\[b_1 z + b_2 z^2 + b_3 z^3+\dots\], 
\end{align}  
where $z=exp(\mu/T)$ is the fugacity for the system with chemical
potential $\mu$. In this expansion, all the density
dependence is in the fugacity $z$. This is a valid expansion for $z\ll
1$ or equivalently for $n \lambda^3\ll 1$ as shown
later in Eq.~\ref{numberdensity}.

The dimensionless virial
coefficients 
$b_n$ depend on the vacuum interaction and the thermal momentum
$\sqrt{MT}$. For a dilute system with thermal wavelength $\lambda$ much
larger than the range $R$, the interaction depends only on the
scattering length $a$.  
Thus, as $a\rightarrow\pm\infty$ at the unitarity limit,
there are no relevant scale left in the interaction and 
the dimensionless coefficients $b_n$ must be \emph{universal}. This is
shown explicitly with a calculation up to the third order in the virial 
expansion. There would be small corrections to these universal results 
from effective range, higher-partial
waves, etc., that are neglected in this calculation but are 
otherwise straightforward to incorporate in perturbation. 

The virial coefficient $b_n$ receives contribution from up to and
including $n$-body physics~\cite{Huang}. For example, $b_4$ would receive
contributions from $1$-body (non-interacting theory), $2$-,$3$- and
$4$-body physics. $b_2$ is the first coefficient that receives
contribution from the interacting theory and it is related to the
2-body scattering phase shift~\cite{BethUhlenbeck,Huang,BedaqueRupak}: 
\begin{align}\label{virial_b22}
b_2^{(2)}&= \frac{1}{\sqrt{2}} e^{\frac{\gamma^2}{MT}}
\[1+ \mathrm{Erf}\(\frac{\gamma}{\sqrt{MT}}\)\],\\
\mathrm{Erf}(x)&= \frac{2}{\sqrt{\pi}}\int_0^x dy e^{-y^2},\nonumber 
\end{align}
where $\gamma =1/a$.
We use the following
notation:
\begin{align}
b_n &= b_n^{(1)}+b_n^{(2)}+\dots+b_n^{(n-1)}+b_n^{(n)},\\
\bar b_n &= b_n-b_n^{(1)},\nonumber
\end{align} 
where
$b_n^{(l)}$ is the $n$-th virial coefficient from $l$-body
contribution. The thermodynamic pressure $P_k^{(l)}$, number density
$n_k^{(l)}$, 
etc., are also denoted in a similar manner. We also define 
$P^{(l)}$ and $n^{(l)}$ without a subscript
 to denote the contribution to the pressure and number density from 
$l$-body physics to all orders in the $z$ expansion.

\begin{figure}[thb]
\includegraphics[width=8.2cm]{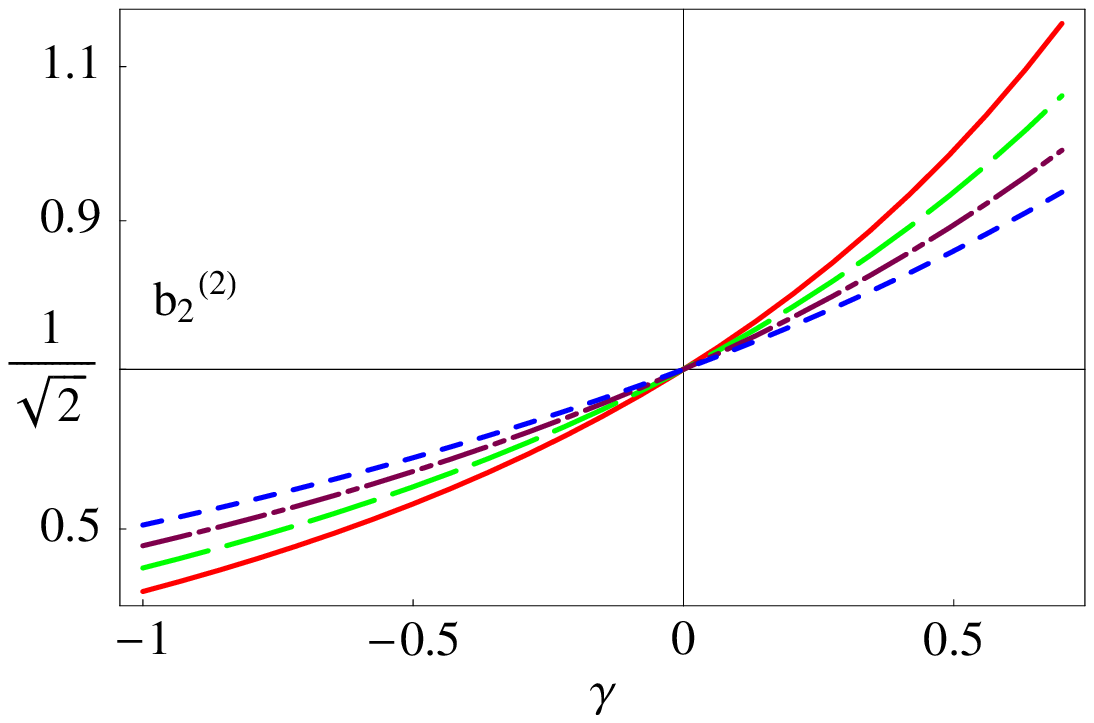}
\includegraphics[width=8cm]{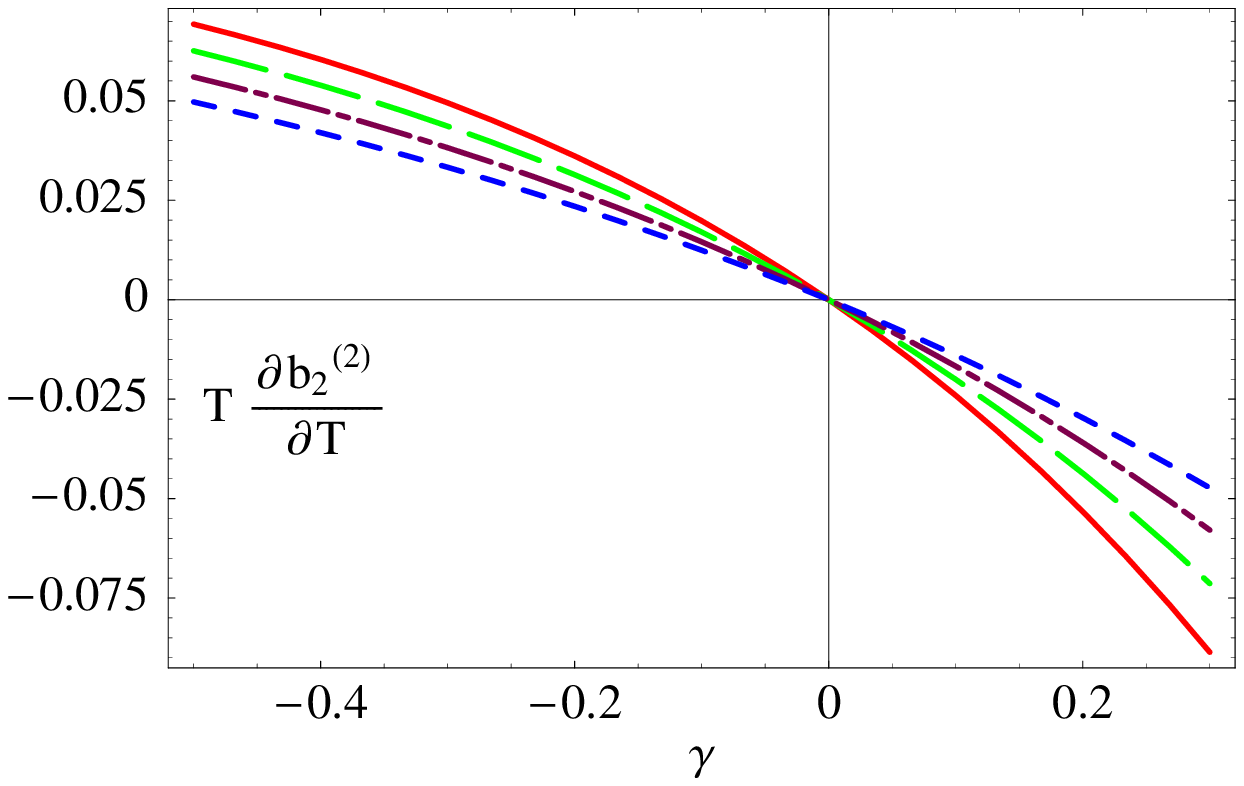}
\caption{\protect Plot of $b_2^{(2)}$, the 2-body contribution to the second
  virial coefficient and its derivative 
$T\partial_T b_2^{(2)}$ for $^6$Li 
as a function of the inverse scattering length 
$\gamma=1/a$ in eV. $1$eV$\approx 5.06\times 10^{-4} \mathring{A}^{-1}$
  in natural units ($\hbar =1 = c$). 
Solid, long-dashed, dot-dashed and short-dashed curves 
correspond to
  temperatures of $6\mu$K, $8.4\mu$K, $11.76\mu$K and $16.46\mu$K
  respectively (in
  $40\%$ increments). The second virial coefficient is temperature
  independent and universal as $\gamma=1/a\rightarrow 0^\pm$.}
\label{figb22}
\end{figure}

From Eq.~\ref{virial_b22}, $b_2^{(2)} =1/\sqrt{2}$ as 
$a\rightarrow\pm\infty$ and it is universal. A plot of $b_2^{(2)}$ and
$T \partial b_2^{(2)}/\partial T$ is
shown in Fig.~\ref{figb22} near the unitarity limit for
lithium $^6$Li for temperatures around $6\mu$K. 

Traditionally the higher
virial coefficients are calculated in a cluster expansion~\cite{BethUhlenbeck} 
and there is
no direct simple formula for relating it to many-body interaction. 
For example, it is not clear how the contribution from 
3-body bound states associated with Efimov effect~\cite{Efimov} 
in bosons such as $^4$He near $a\rightarrow\pm\infty$ is taken into account. 
This is in contrast to an effective field theory calculation where the
virial coefficients are directly related to many-body scattering and
the $^4$He 3-body 
bound state contribution can be directly taken into account. The
general methodology for such an effective field theory calculation was
developed earlier in Ref.~\cite{BedaqueRupak}. 

%=====================================================================
%\end{section}
%\begin{section}{Calculation}
%====================================================================
For a dilute system with a clear separation of scales $1/|a|\ll
n^{1/3}\ll\sqrt{MT/(2\pi)}\ll1/R$, effective field theory
is quite ideal, and provides model independent results in the region of
interest. 
A non-relativistic system near the Feshbach resonance 
in the $S$-wave at chemical potential $\mu$ (or densities $n$) 
and temperature $T$ such that 
 $\sqrt{M\mu}$ (or $n^{1/3}$), $\sqrt{MT}\ll 1/R$ can be described by the 
Lagrangian density:
\begin{align}\label{Lagrangian}
\mcal L = \psi^\dagger\(i\partial_0 +\frac{\nabla^2}{2 M}\)\psi
+ \mu\ \psi^\dagger \psi
-\frac{g}{4}\(\psi\sigma_2\psi\)^\dagger\(\psi\sigma_2\psi\)+\dots ,
\end{align} 
where the ``$\dots$'' represents higher dimensional operators that are
suppressed for dilute systems, $\psi$ are the spin-$\frac{1}{2}$
fermion fields and the $\sigma_i$ matrices act in the spin space. The 
four-fermion coupling $g$ is related to the  2-body scattering
length $a$ (see Refs.~\cite{CRS,BvK} and references therein):
\begin{align}
g(\nu)&=-\frac{4\pi}{M}\frac{1}{\nu-1/a},
\end{align}
where $\nu$ is the renormalization scale in dimensional
regularization. One can choose to regulate using momentum cutoff
$\Lambda$ and get a $g(\Lambda)$ similar in form to $g(\nu)$ 
above~\cite{BedaqueRupak}. The final result is independent of the
regularization scheme and renormalization 
scale as it should be. Notice that there
are no 3-body $S$-wave operator at leading
order~\cite{BvK,BRHG}. Therefore when we calculate the third virial
coefficient, it will depend only on the 2-body interaction and will be
universal as the 2-body scattering length $a$ approaches $\pm\infty$.

The organization of the fugacity $z$ expansion in terms of Feynman diagrams 
was worked out earlier in 
Ref.~\cite{BedaqueRupak} and applied to bosons. For bosons, there is a
3-body interaction at leading order which can lead to a 3-body bound state that
dominates the third virial coefficient $b_3$. 
At the same time this prevents $b_3$ from being universal for bosons since the
3-body binding energy is system dependent and 
does not necessarily approach an universal value as
$a\rightarrow\pm\infty$. 

\begin{figure}[thb]
\includegraphics[width=12cm]{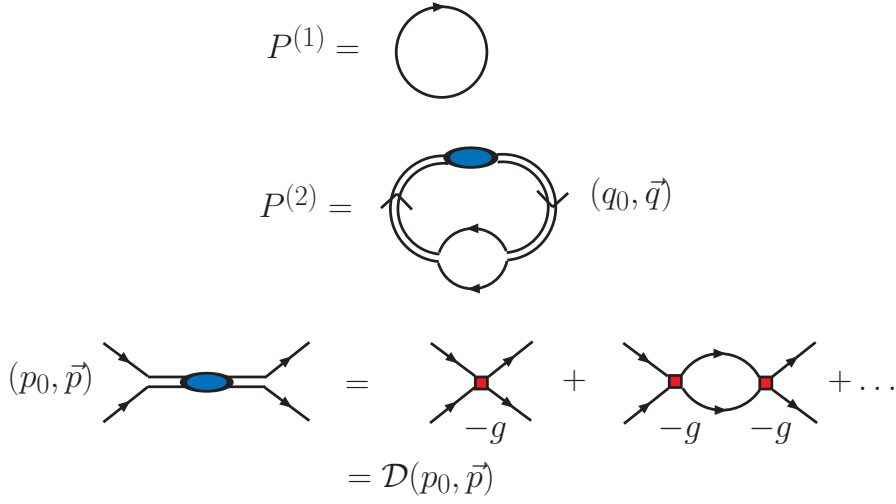}
\caption{\protect The pressure $P$ from 1- and 2-body physics. Solid lines are
  fermion fields. Each particle-particle loop is order $g\sqrt{M
  T}\sim a\sqrt{MT}$ and summed to all orders to form the dressed dimer
  propagator, represented by the double lines. }
\label{fig_p12}
\end{figure}

The first two virial coefficients $b_1$ and $b_2$ are
known. Nevertheless, it is
instructive to derive them in the effective field theory from Feynman
diagrams before
calculating the third virial coefficient $b_3$. Feynman diagrams with
a closed particle loop is order $z$ (from the energy integral over the 
Matsubara frequency) and vanish in vacuum as expected. 
This can be seen by calculating 
the pressure in the free theory which is given by the loop diagram shown
in Fig.~\ref{fig_p12}:
\begin{align}\label{1body_pressure}
P^{(1)} &= 2 T\sum_{n=-\infty}^{\infty}\int\frac{d^3\ q}{(2\pi)^3}
\log\[i(2n+1)\pi T+\mu -\frac{{\bm q}^2}{2M}\] = 
2 T \int\frac{d^3\ q}{(2\pi)^3}\log\[1+z \exp\(\frac{\bm q^2}{2 MT}\)\]\\
&=\frac{2 T}{\lambda^3}\[z-\frac{z^2}{4\sqrt{2}}
+\frac{z^3}{9\sqrt{3}}-+\dots\], \nonumber\\
\Rightarrow b_1&+ b_2^{(1)}+b_3^{(1)}+\dots
=1-\frac{1}{2^{5/2}}+\frac{1}{3^{5/2}}-+\cdots .\nonumber 
\end{align}
A closed particle-particle diagram (only possible with interactions) is 
$\mcal O(1)$ and does not vanish in the vacuum. However, a closed loop
with ``baryon number'' equal to 2 i.e. a closed dimer propagator 
is order $\mcal O(z^2)$. Similarly a closed trimer propagator is $\mcal
O(z^3)$. This is more easily demonstrated with the following
calculation of the second-virial coefficient.  

At large scattering length $|a|\sqrt{MT/(2\pi)}\gtrsim 1$, 
the dimer propagator is given
by an infinite geometric sum shown in Fig.~\ref{fig_p12}:
\begin{align}\label{dimer_prop}
\mcal D(p_0,\vec{\bm p})&\approx 
\frac{4\pi}{M }
\frac{1}{-\frac{1}{a}+\sqrt{\frac{ p^2}{4} -M p_0 -2 M\mu}}
+\mcal O(z^2),
\end{align}  
where $p=|\vec{\bm p}|$, $q=|\vec{\bm q}|$ and 
$p_0$ is an odd multiple of $i\pi T$. 
Some care is needed in evaluating $\mcal D(p_0,\vec{\bm p})$ where
$p_0$ is to be identified with an ``final'' loop integral variable
in Eqs.~\ref{P_2} and~\ref{3body_density}.  
In the
center of mass frame $\vec{\bm p}=\vec{\bm k} -\vec{\bm k} = 0$, 
analytically continuing
$p_0$ to the center 
of mass energy in vacuum $p_0\rightarrow k^2/M-2\mu +i0^+$, 
the dimer propagator $\mcal D(
k^2/M-2\mu,0)$ reproduces the correct large scattering length 2-body
amplitude: 
\begin{align}
\mcal A(k) = \frac{4\pi}{M}\frac{1}{-\frac{1}{a}-i k} .
\end{align} 
The contribution from
the dimer propagator to the pressure is
\begin{align}\label{P_2}
P^{(2)}&= -T\sum_{n=-\infty}^\infty\int\frac{d^3\ q}{(2\pi)^3}
\log\[\mcal D(i(2n+1)\pi T,\vec{\bm q})\]\\ 
&\approx-\frac{1}{2\pi i}\oint \frac{d\eta}{exp\(\eta/T\)+1}
\int\frac{d^3\ q}{(2\pi)^3}\[
\log \(-\frac{1}{a}+\sqrt{\frac{q^2}{4}- M\eta -2 M\mu}\)
+\mcal O(z^2)
\]\nonumber.
\end{align}
The contour integral in $\eta$ is in an anti-clockwise sense over all
the odd integer multiples of $i\pi T$. Looking at 
Eqs.~\ref{dimer_prop} and \ref{P_2} , one can see how the energy integral
over $\eta$ for a dimer propagator are $\mcal O(z^2)$ because a dimer 
carries chemical potential $2\mu$. This is similar 
to the energy $\eta$ integral for a closed single particle propagator
that carries chemical potential $\mu$ and 
$1/[\exp(\beta\eta)+1]\sim\mcal O(z)$ for $\eta\sim-\mu$.  
$P^{(2)}$ is $\mcal O(z^2)$ and an explicit calculation shows:
\begin{align}\label{2body_pressure}
P^{(2)}&\approx P_2^{(2)}+ P_3^{(2)} +\mcal O(z^4),\\
P_2^{(2)}&=\frac{\sqrt{2} T}{\lambda^3}z^2 
e^{\frac{\gamma^2}{MT}}
\[1+ \mathrm{Erf}\(\frac{\gamma}{\sqrt{MT}}\)\],\nonumber\\
P_3^{(2)}&=0 \Rightarrow b_3^{(2)}=0.\nonumber
\end{align} 
Comparing Eqs.~\ref{virial_exp}, \ref{virial_b22}, \ref{1body_pressure} and
\ref{2body_pressure}, we see that the known results up
to the second order in the virial expansion are exactly 
reproduced in the effective field
theory calculation.

\begin{figure}[thb]
\includegraphics[width=16cm]{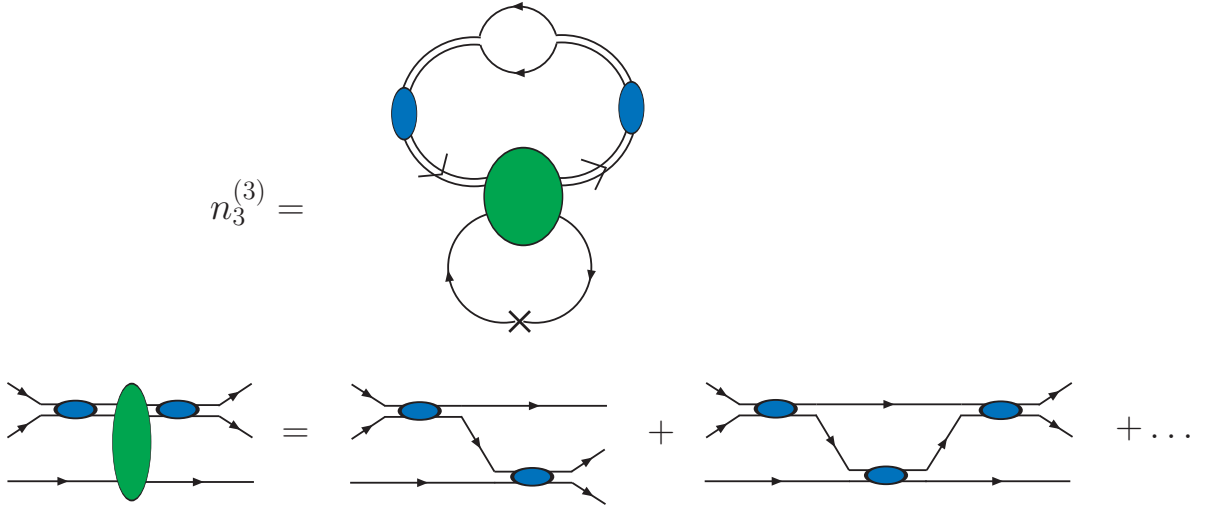}
\caption{3-body contribution to the number density $n_3^{(3)}$.}
\label{fig_n3}
\end{figure}
The calculation of the third virial coefficient $b_3^{(3)}$ is technically more
challenging but it follows the same procedure. 
First, the leading order trimer propagator in the fugacity $z$ expansion
is calculated, and then the energy integral over imaginary odd
multiples of $i\pi T$ is carried out.   
The trimer propagator is $\mcal O(1)$ in the $z$ expansion 
and carries chemical potential $3\mu$. Thus an energy 
integration over a closed trimer loop contributes at $\mcal O(z^3)$. 
Like the dimer, 
the trimer propagator requires summation of an infinite series 
of Feynman diagrams 
for 
$|a|\sqrt{MT/(2\pi)}\gtrsim 1$. However, unlike the dimer, the 3-body
contributions do not form a geometric series and this makes the trimer
more challenging. The solution to the trimer problem has been 
know~\cite{BvK,BRHG} and
we will not repeat it here. The relevant Feynman diagrams are shown in 
Fig.~\ref{fig_n3}. It is easier to calculate the number density 
$n=\partial_\mu P$ instead of the pressure $P$ for the trimer. We find:
\begin{align}\label{3body_density}
n_3^{(3)} =&-i\frac{9\sqrt{3}M}{4\pi^2\lambda^3}z^3
\oint d\eta\int_0^\infty dk\ k^2
\frac{\gamma+\sqrt{\frac{3}{4}k^2-M\eta}}
{\sqrt{\frac{3}{4}k^2-M\eta}\(\frac{3}{4}k^2-M\eta-\gamma^2\)^2}
e^{-\eta/T} a(\eta,k,k)\\
&{}-i\frac{9\sqrt{3}M}{2\pi^2\lambda^3}z^3
\oint d\eta\int_0^\infty dk \frac{k^2}
{\frac{3}{4}k^2-M\eta-\gamma^2}e^{-\eta/T}b(\eta,k,k)\nonumber\\
=& \partial_\mu P_3^{(3)}= \frac{6}{\lambda^3}
b_3^{(3)} z^3,\nonumber
\end{align} 
where the contribution from the trimer propagator, after projecting
onto the $S$-wave, is defined through 
the integral equations
\begin{align}
a(\eta, k, k')=& -\frac{1}{2}K(\eta, k, k')
-\frac{1}{\pi}\int_0^\infty dl \frac{l^2 a(\eta,k,l)}
{l^2-\frac{4}{3}(M\eta+\gamma^2)} K(\eta, l, k'),\\
b(\eta, k, k')=&-\frac{8}{3}\frac{\gamma+\sqrt{\frac{3}{4}k^2-M\eta}}{
(k^2+k'^2+k k'-M\eta)(k^2+k'^2-k k'-M\eta)}\nonumber\\
&-\frac{2}{\pi}\int_0^\infty dl \frac{l^2 b(\eta,k,l)}
{l^2-\frac{4}{3}(M\eta+\gamma^2)} K(\eta, l, k'),\nonumber\\
K(\eta,k,k')=&\frac{4}{3}\frac{\gamma+\sqrt{\frac{3}{4}k^2-M\eta}}{k
  k'}
\log\[\frac{k^2+k'^2+k k'-M\eta}{k^2+k'^2-k k'-M\eta}\].\nonumber
\end{align}

%=====================================================================
%\end{section}
%\begin{section}{Analysis}
%====================================================================
 \begin{figure}[thb]
\includegraphics[width=8.2cm]{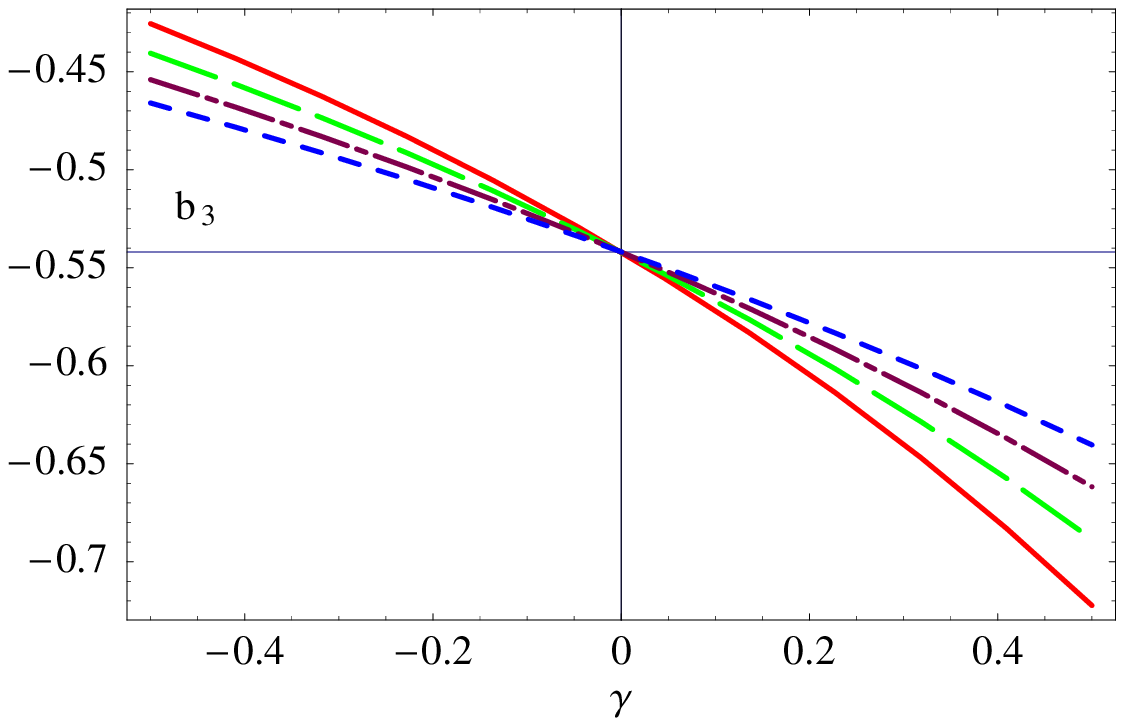}
\includegraphics[width=8cm]{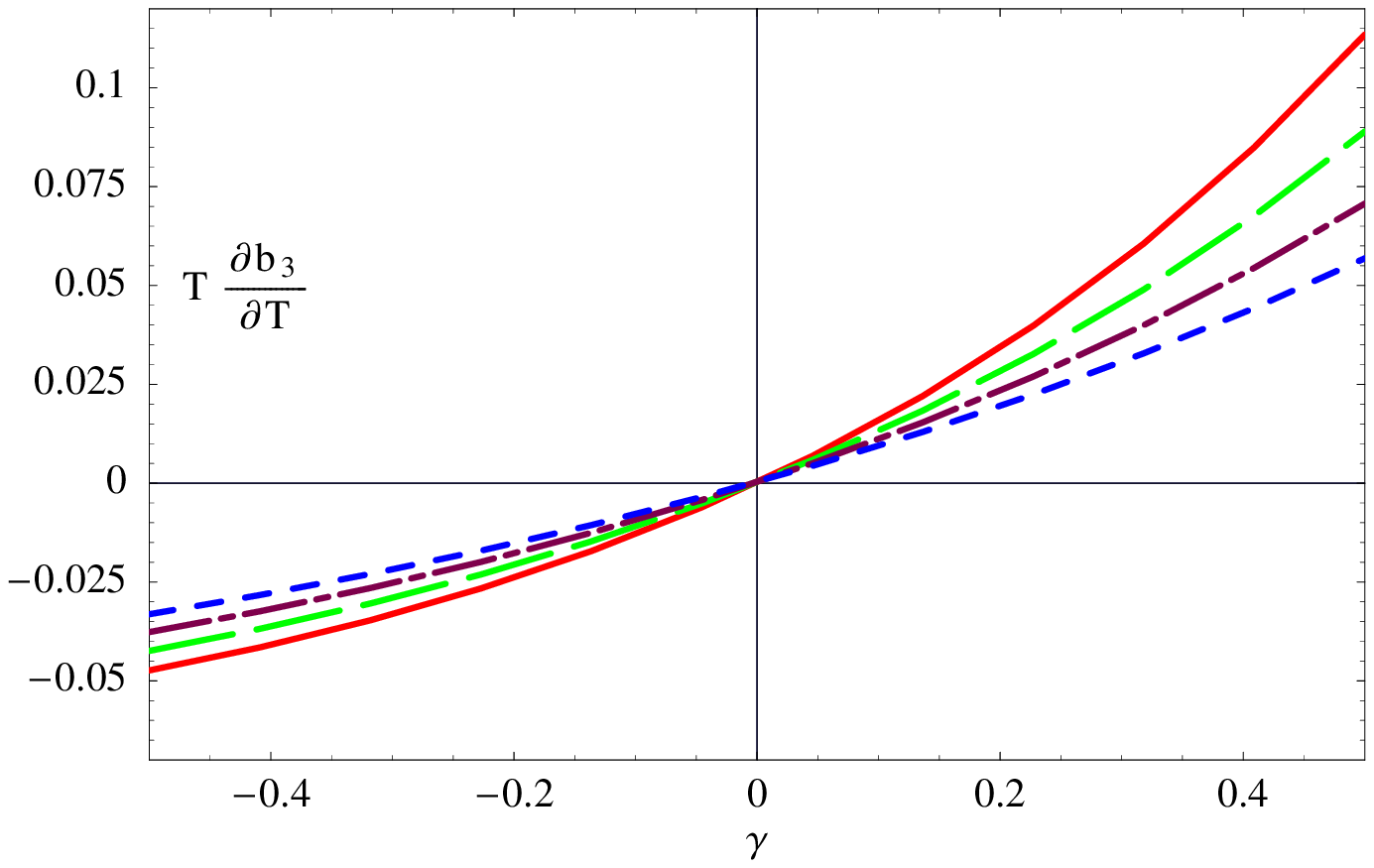}
\caption{\protect Plot of $\bar b_3$ and its derivative $T\partial_T \bar b_3$
  for $^6$Li as a function of $\gamma =1/a$ in eV. 
The different curves are for the same set of temperatures as in
  Fig.~\ref{figb22} and uses the same notations. $\bar b_3$ shows
  universal behavior.}
\label{figb3}
\end{figure}
In Fig.~\ref{figb3}, the third virial coefficient 
$\bar b_3= b_3 - b_3^{(1)}=b_3^{(2)}+b_3^{(3)}=b_3^{(3)}$ and its derivative 
$T\partial_T\bar b_3$ are shown for $^6$Li at temperatures 
around $6\mu$K. As expected from physical arguments presented earlier, 
the third virial coefficient is \emph{universal}. This can be made
more explicit in Eqs.~\ref{2body_pressure} and \ref{3body_density} 
for $|a|\rightarrow\infty$, and write, for example, 
the coefficient $b_3^{(3)}$ in terms of dimensionless integrals.
We find $\bar b_3^\infty\approx-0.54$.

From this calculation, we see that the virial coefficients up to
$\mcal O(z^3)$ are not un-naturally large [numbers of $\mcal O(1)$]
and for $z\ll1$, the virial
expansion seems consistent. 
Formally, the expansion in $z$ is
equivalent to an expansion in diluteness parameter in 
perturbation \cite{Huang,HoMueller} for 
$n\lambda^3\ll 1$:
\begin{align}\label{numberdensity}
n &= \frac{\partial P}{\partial\mu} 
\approx \frac{2}{\lambda^3}\[b_1 z +2 b_2 z^2 + 3 b_3 z^3\]+\dots,\\
\Rightarrow z &\approx \frac{n\lambda^3}{2}
-2b_2\( \frac{n\lambda^3}{2}\)^2+(8b_2^2-3b_3)\(
\frac{n\lambda^3}{2}\)^3+
\dots,\nonumber
\end{align}
where we used $b_1=1$. 
Defining
the density $n$ in terms of the Fermi temperature $T_F$, 
$n\lambda^3/2 = 4/(3\sqrt{\pi}) (T_F/T)^{3/2}$ is the
small expansion parameter for $T\gg T_F$. The energy density can be
obtained from the pressure using standard thermodynamic relations 
$\epsilon = -P +\mu\ n +T\ s$, where $s=\partial_T P$ is the entropy
density.  

As $a\rightarrow\pm\infty$, the energy density
can be written as~\cite{HoMueller}:
\begin{align}\label{universal_energy}
\epsilon^\infty&= \frac{3}{2}T n\[1+\frac{1}{2^{5/2}}\frac{n\lambda^3}{2}+
\frac{n^2\lambda^6}{4}\(\frac{1}{8} -\frac{2}{9\sqrt{3}}\)\]
+\epsilon^\infty_\mathrm{int}\equiv \epsilon_\mathrm{kin}
+\epsilon^\infty_\mathrm{int},\\
\epsilon^\infty_\mathrm{int}&= \frac{3}{2}T n\frac{n\lambda^3}{2}
\[-b_2^{\infty(2)}+\frac{n\lambda^3}{2}
\(-\sqrt{2}b_2^{\infty(2)}+4 (b_2^{\infty(2)})^2-2 \bar
b_3^{\infty}\)\].
\nonumber
\end{align}
$\epsilon^\infty_\mathrm{int}$ is the contribution from the
interacting theory in Eq.~\ref{Lagrangian} in the unitarity limit 
$a\rightarrow\pm\infty$.

\begin{figure}[thb]
\includegraphics[width=8.2cm]{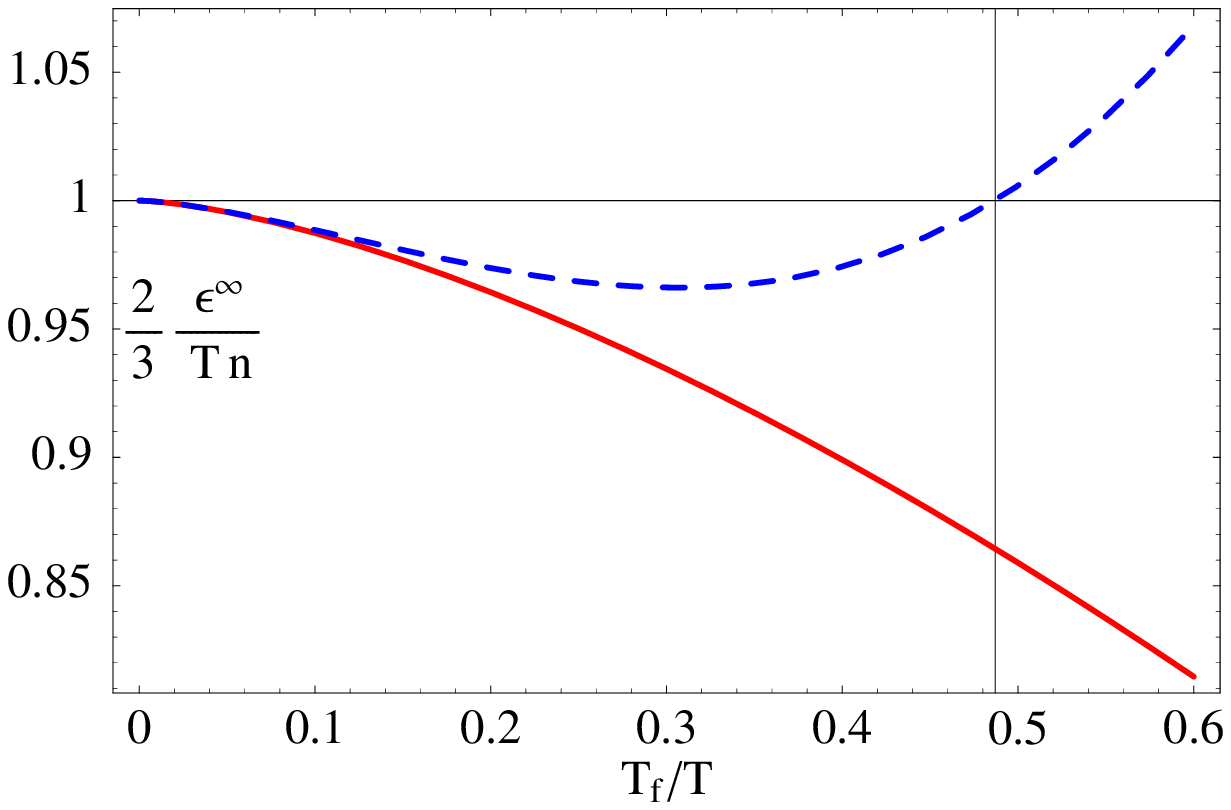}
\includegraphics[width=8cm]{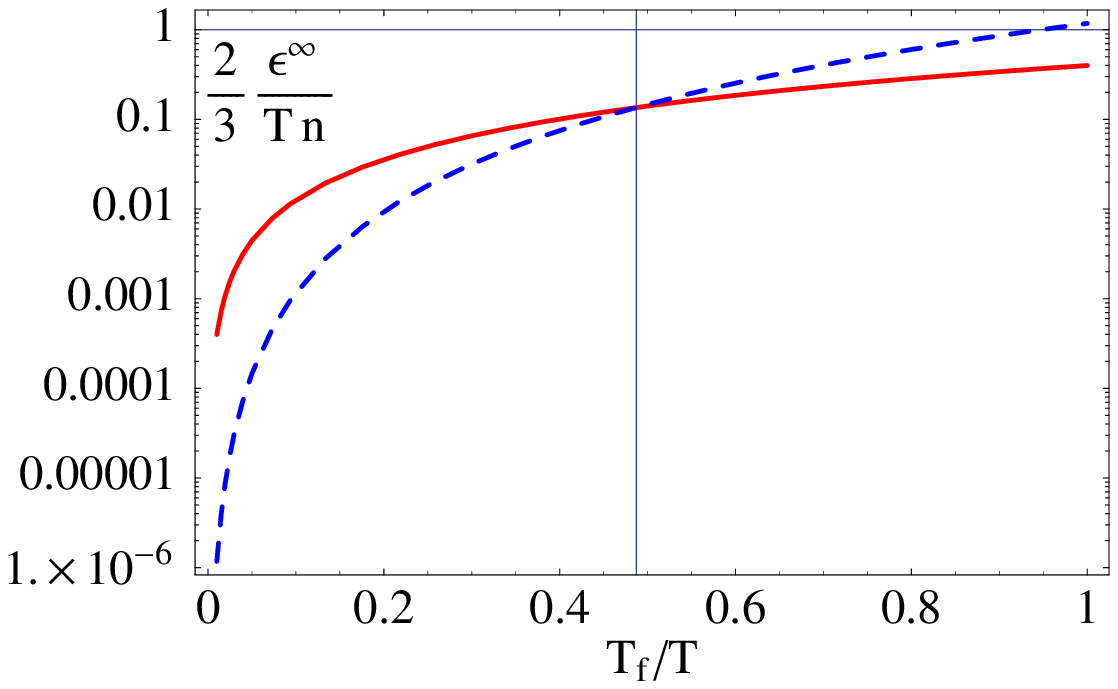}
\caption{\protect Left subplot: the second (solid curve) and third order
  (dashed curve) energy density $\epsilon^\infty$ 
in the $n\lambda^3/2$ expansion in
  Eq.~\ref{universal_energy} as a fraction of the leading order result $3 T
  n/2$ vs. $T_F/T$.   
Right subplot: Convergence (error) plots~\cite{Lepage}. 
Absolute value of the ratio of only the 
$\mcal O(n\lambda^3)$ to the leading order contribution  
  (solid curve) and only the $\mcal O(n^2\lambda^6)$ to the leading order 
contribution (dashed curve) 
to the energy density as a function of $T_F/T$. For small
  values of $T_F/T$, the $\mcal O(n^2\lambda^6)$ effects are smaller than 
$\mcal O(n\lambda^3)$. These effects are equal at around $T\sim 2 T_F$
  and the perturbation breaks down. Note the $\mcal O(n^2\lambda^6)$
  effects equal the leading order result (ratio equals $1$) at a
  smaller temperature $T\sim 1 T_F$, at fixed $T_F$. 
The $\mcal O(n\lambda^3)$ results are smaller than
  the leading order results to even smaller temperatures (not shown).      
}
\label{energydensity}
\end{figure}

In Fig.~\ref{energydensity}, we have plotted the contributions from the
third virial coefficient to the energy density. 
In the left subplot in Fig.~\ref{energydensity}, the solid curve shows the
contribution up to $\mcal O(n\lambda^3)$ normalized to the leading
order contribution as a function of $T_F/T$. The dashed curve is the
$\mcal O(n^2\lambda^6)$ contribution which is a smaller correction for 
$T_F/T \lesssim 0.49$. 
This is shown more clearly in the right subplot in
Fig.~\ref{energydensity}: the convergence (error) plot~\cite{Lepage}. 
It is deduced that the range of convergence is $T_F/T\lesssim 0.49$. 

\begin{figure}[thb]
\includegraphics[width=8cm]{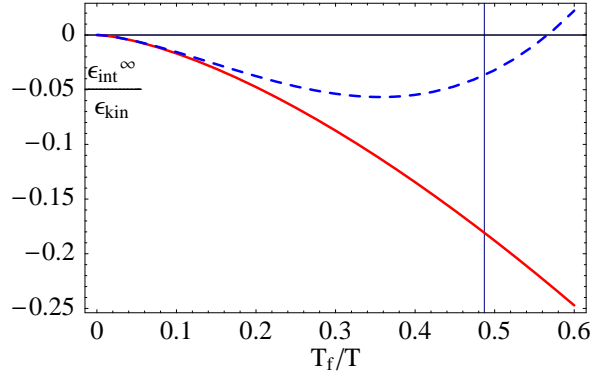}
\caption{\protect Ratio $\epsilon^\infty_\mathrm{int}/\epsilon_{kin}$
of interaction to kinetic energy
   as
a function of $T_F/T$ in the range of convergence $T_F/T\lesssim 0.49$
deduced in Fig.~\ref{energydensity}. Solid curve: leading order 
$[\mcal O(n\lambda^3)]$ 
result, dashed curve: next-to-leading order $[\mcal O(n^2\lambda^6)]$ result. }
\label{energyratio}
\end{figure}

The ratio of the interaction energy $\epsilon_\mathrm{int}$ to the
contribution from the free Lagrangian density $\epsilon_\mathrm{kin}$ 
is shown in 
Fig.~\ref{energyratio} for a range of $T_F/T$ where the virial
expansion converges at this order of the calculation.  For the
experiments on $^6$Li, at temperatures around $T\sim 3 T_F$ we expect
the virial expansion to be valid. At this temperatures, the effect of
the third virial coefficient should be detectable in experimental data
for dilute systems satisfying the constraint 
$n^{1/3}\ll\sqrt{MT/(2\pi)}\ll 1/R$. 

%======================================================================
%\end{section}
%\begin{section}{Summury}
%=====================================================================
In conclusion, we considered a spin-$\frac{1}{2}$ dilute fermionic
system near the unitarity limit $|a|\rightarrow\infty$. At finite
temperature the thermodynamic pressure was calculated in a model
independent way using effective field theory up to the third order in the
virial expansion. At this order, universality was demonstrated and
a range of temperature $T\gtrsim 3 T_F$ for a fixed density (conversely a range of
densities for a fixed temperature) was identified for a
self-consistent application of the results.

The virial expansion is an useful tool for dilute systems where the
microscopic physics is poorly known or difficult to calculate. 
The universal
forms for the equation of state derived here are applicable to a wide class of
physical problems in nuclear and atomic physics. These results are relevant 
for understanding neutrino physics in hot    
dilute nuclear matter $n\sim 10^{-4}$ fm$^3$~\cite{HorowitzSchwenk}, 
and the same equations for
energy density, etc., are testable in atomic physics experiments. At
lower temperatures, numerical lattice calculations are more
appropriate~\cite{Wingate:2005xy,Bulgac:2005pj,Lee:2005is}. 
However, at intermediate temperatures, our results
could be used as checks for the lattice calculations.   

The author thanks P. Bedaque and A. Schwenk for a critical reading of
the manuscript. 
This work was supported in part by DOE grants DE-FG02-00ER41132 
and DE-FC02-01ER41187 .
%=======================================================================
%\end{section}
%=======================================================================
\bibliographystyle{unsrt}
%\bibliographystyle{h-physrev4}
%\bibliography{bibFermiVirial}

\end{document}